\documentclass[format=acmsmall, review=false]{acmart}
\usepackage{acm-ec-25}
\usepackage{booktabs} 
\usepackage[ruled]{algorithm2e} 

\SetAlFnt{\small}
\SetAlCapFnt{\small}
\SetAlCapNameFnt{\small}
\SetAlCapHSkip{0pt}
\IncMargin{-\parindent}
\usepackage{subcaption}
\usepackage{array}

\setcitestyle{authoryear}

\title[]{Beyond Revenue and Welfare: Counterfactual Analysis of Spectrum Auctions with Application to Canada's 3800MHz Allocation}

\author{Sara Jalili Shani, Kris Joseph, Michael B. McNally, James R. Wright}

\begin{abstract}
Spectrum auctions are the primary mechanism through which governments allocate scarce radio frequencies, with outcomes that shape competition, coverage, and innovation in telecommunications markets. While traditional models of spectrum auctions often rely on strong equilibrium assumptions, we take a more parsimonious approach by modeling bidders as myopic and straightforward: in each round, firms simply demand the bundle that maximizes their utility given current prices. Despite its simplicity, this model proves effective in predicting the outcomes of Canada’s 2023 auction of 3800 MHz spectrum licenses. Using detailed round-by-round bidding data, we estimate bidders’ valuations through a linear programming framework and validate that our model reproduces key features of the observed allocation and price evolution. We then use these estimated valuations to simulate a counterfactual auction under an alternative mechanism that incentivizes deployment in rural and remote regions, aligning with one of the key objectives set out in the Canadian Telecommunications Act. The results show that the proposed mechanism substantially improves population coverage in underserved areas. These findings demonstrate that a behavioral model with minimal assumptions is sufficient to generate reliable counterfactual predictions, making it a practical tool for policymakers to evaluate how alternative auction designs may influence future outcomes. In particular, our study demonstrates a method for counterfactual mechanism design, providing a framework to evaluate how alternative auction rules could advance policy goals such as equitable deployment across Canada.
\end{abstract}

\begin{document}

\begin{titlepage}

\maketitle

\vspace{1cm}
\setcounter{tocdepth}{2} 
\tableofcontents

\end{titlepage}
\section{Introduction} \label{sec: introduction}

Spectrum auctions play a crucial role in allocating limited radio frequency resources to telecommunication companies. These auctions enable governments to efficiently distribute spectrum licenses, thereby promoting competition and innovation in the industry \citep{CramtonKwerelRosstonSkrzypacz2011}. The revenue generated from spectrum auctions can be substantial, often contributing significantly to government budgets and funding various public initiatives \citep{KwerelRosston2000Insiders, Taylor2013v38n1a2600}. For this reason, changes in auction policy cannot be made arbitrarily, as both fiscal outcomes and long-term market structure are at stake. Nevertheless, regulatory interventions are often necessary to ensure that market outcomes align with broader public objectives rather than pure revenue maximization.

According to the Canadian Radio-television and Telecommunications Commission (CRTC), the promotion of rural broadband coverage is one the key priorities of their departmental plan for the year 2025-26 \citep{crtc2025dpge}. This has previously been an objective of the federal government, and they have pursued a mix of direct investment programs and regulatory measures to encourage deployment in underserved areas since 2016 \citep{ISED2022Consultation, ised2019_connectivity}. However, despite these efforts, the CRTC’s \emph{Communications Monitoring Report 2025} indicates that by the end of 2023, although 93.3\% of the national population had access to 5G networks, coverage in First Nations reserves and rural areas stood at only 47\% and 71.2\%, respectively \citep{crtc2025policy}. These disparities suggest that existing initiatives remain insufficient to achieve the coverage goals articulated by the CRTC.

A complementary policy avenue is to explore whether auction design itself can influence deployment outcomes. Advancing this idea requires a simulation or predictive tool capable of evaluating, ex ante, how modifications to auction rules may affect both allocative efficiency and coverage. Such a tool would provide regulators with a systematic framework for policy design and for examining counterfactual scenarios within spectrum allocation processes. Performing counterfactual analysis in complex spectrum auctions, however, is far from straightforward. Spectrum licenses are typically purchased as bundles, and their values exhibit strong complementarities and substitution effects. Two adjacent geographic licenses or contiguous frequency blocks, for instance, may be far more valuable together than separately \citep{ConnollyXiaoTanLim2021}. To account for these interdependencies, regulators have adopted iterative auction formats such as the Simultaneous Multiple Round Auction (SMRA) and the Combinatorial Clock Auction (CCA) \citep{Milgrom2000, AusubelCramtonMilgrom2006}. These mechanisms unfold across multiple rounds of bidding, allowing bidders to adjust their demand as prices rise. The resulting environment is strategically intricate, involving intertemporal decision-making, combinatorial valuation, and dynamic competition across hundreds of licenses.

From a methodological standpoint, there are two main research paradigms for conducting counterfactual analysis in such auctions.  
\paragraph{Equilibrium-Based Approaches.}
The first relies on \emph{equilibrium analysis}, where one models the auction as a game of incomplete information under hypothetical preferences, and computes its equilibrium outcomes. The idea is to predict how rational bidders would behave under a new rule, by calculating the equilibrium under the alternative design. Yet, for realistic combinatorial or clock auctions, finding equilibrium strategies is computationally intractable \citep{BichlerShabalinZiegler2013, GretschkoKnapekWambach2016}. The strategy space grows exponentially with the number of licenses, and the multiplicity of equilibria makes outcome prediction ambiguous \citep{PalaciosHuertaParkesSteinberg2024}.  

Recent research has increasingly turned toward approximation and learning-based equilibrium computation. A variety of machine-learning and algorithmic approaches have been proposed to approximate equilibria in clock auctions, with neural-network methods and reinforcement learning playing central roles. Using gradient dynamics, \citet{HeidekruegerSuttererKohring2021} introduce \emph{Neural Pseudogradient Ascent}, a scalable multi-agent reinforcement learning method in which bidding strategies are represented by neural networks trained via pseudogradient dynamics in self-play to compute approximate Bayesian Nash equilibria. In another gradient-based approach, \citet{BichlerFichtlOberlechner2022} propose \emph{Simultaneous Online Dual Averaging}, a discretization-based online convex-optimization framework that applies dual averaging with entropy regularization to compute approximate Bayes–Nash equilibria as distributional strategies in continuous-type Bayesian auction games.

Beyond gradient methods, \citet{10591726} extend their earlier \emph{Simultaneous Move Monte Carlo Tree Search} algorithm to an incomplete-information setting, developing bidding strategies that jointly address four key strategic challenges: the exposure problem, own-price effects, budget constraints, and eligibility management. They conduct empirical equilibrium analysis by constructing normal-form games from 1{,}000 simulated instances ($n=3$ bidders, $m=9$ items) and verifying Nash equilibria via best-response deviation checks. In work applying reinforcement learning directly to equilibrium approximation, \citet{greenwald2012approximating} propose a two-phase predict-and-optimize method combining best-reply dynamics with decision-theoretic prediction to compute approximate Bayesian Nash equilibria in sequential sealed-bid auctions with incomplete information and multi-unit demand. Using multi-agent reinforcement learning, \citet{10.1145/3670865.3673644} analyze tractable spectrum-auction games with imperfect information and asymmetry, verifying approximate equilibria via NashConv. Their case study comparing drop-by-player and drop-by-license bid-processing rules demonstrates substantial changes in bidder behavior, revenue, welfare, and auction length relative to myopic heuristics. Similarly, \citet{Pieroth2023EquilibriumCI} use deep reinforcement learning in self-play to compute approximate Bayesian Nash equilibrium strategies in multi-stage games with continuous signals and actions, including sequential auctions and elimination contests. Most recently, \citet{ThomaBosshardSeuken2025} introduce a custom discretization of continuous type and action spaces augmented with public belief states to construct a finite-state Markov decision process solvable by backward induction. Their method computes stage-wise BNEs via iterated best-response dynamics (pattern-search optimization with Monte Carlo integration), and includes a verification step bounding utility loss through a decomposition theorem.

Among these contributions, only \citet{10.1145/3670865.3673644} explicitly apply their methodology to counterfactual analysis. Nevertheless, many of the equilibrium-approximation techniques developed in these works could, in principle, be adapted for counterfactual evaluation even if this is not their stated objective. Despite these advances, learning-based equilibrium models remain limited by computational scalability and are still unable to accommodate realistic spectrum-auction instances. Moreover, equilibrium selection remains a persistent difficulty, as multiple equilibria can coexist and lead to markedly different welfare outcomes.

\paragraph{Behavioral and Data-Driven Approaches.}
The second line of research takes an empirical approach, recovering bidder preferences or strategies directly from observed auction data rather than computing theoretical equilibria. These methods typically rely on structural econometric models or revealed-preference inequalities to infer underlying valuations, which can then be used to simulate policy counterfactuals. In the context of U.S. Federal Communications Commission (FCC) spectrum auctions, \citet{XiaoYuan2022License} analyze Auction 73 using moment inequalities derived from an incomplete model of bidder behavior to estimate license complementarities and evaluate package-bidding policies. However, the restrictive behavioral assumptions required for tractability limit the predictive performance of their estimates. A related study, \citet{BajariFox2005Complementarities}, applies a semiparametric two-sided matching estimator based on revealed-preference inequalities and pairwise stability, identifying strong geographic scope complementarities. Although they do not conduct counterfactual simulations, their results provide policy insights by suggesting that larger geographic license areas (e.g., 1/12th of the U.S. or European-style nationwide regions) would better capture scope efficiencies and reduce risks of collusion or excessive fragmentation. Building on this line of inquiry, subsequent work measures both within-auction complementarities (across licenses sold in the same event) and cross-auction complementarities (between auctioned and pre-existing holdings), yielding implications for future FCC auction design \citep{ConnollyXiaoTanLim2021}. More recently, \citet{newman2024incentive} conduct more than 1,300 national-scale computational simulations of the FCC’s 2016–2017 incentive auction reverse descending clock, using realistic bidder value models and truthful bidding to perform a postmortem counterfactual analysis. Their study evaluates the impacts of several design elements, including custom repacking feasibility checkers, multistage clearing, VHF repacking, and alternative price trajectories, and demonstrates substantial quantitative effects on procurement costs, efficiency (value loss), and the amount of spectrum cleared. It is important to note, however, that the auction format and direction analyzed in their work differ from those considered here. Moreover, while both approaches leverage revealed-preference constraints under myopic bidding to inform valuation estimates, \citet{newman2024incentive} generate station-specific valuations probabilistically using two aggregate models (MCS from the literature and BD derived from bid bounds) and we estimate per-bidder valuations from round-by-round bids using linear programming.

For Canada’s spectrum auctions, \citet{Beltrán2016Strategic} employ a revealed-preference framework to test whether bidder behavior is consistent with utility maximization. They find that bidders frequently deviate from straightforward bidding and draw on innovations from the CCA to address well-known limitations of the Simultaneous Ascending Auction (SAA). However, their analysis remains primarily descriptive, without conducting bidder-specific structural estimation of valuations or complementarities, and offers no counterfactual simulations, revenue analysis, or design recommendations beyond highlighting CCA improvements over the SAA.

\paragraph{Our Approach.}
The main goal of this paper is to evaluate how the outcomes of an auction with the same participants as Canada’s 3800 MHz auction would change under alternative design choices. To do so, we must first simulate the original 3800 MHz auction before assessing the effects of modified policy rules. Addressing this first step requires estimating bidders’ valuations from the observed round-by-round bidding data, as these valuations are not directly observable. Given the strategic and computational complexity of spectrum auction formats such as the SMRA and the CCA, we introduce simplifying behavioral assumptions that preserve key strategic features of the original environment while ensuring tractability.

Bidders are assumed to behave \emph{myopically}: in each round, they select the bundle that maximizes their immediate utility given the current posted prices, without anticipating future rounds or strategic reactions. Each bidder’s feasible set of bundles is restricted to those actually submitted during the real auction, reflecting empirical findings that spectrum bidders tend to pre-select a limited set of preferred package combinations before bidding begins  \cite{Scheffel2012}. 

Each bidder’s valuation is modeled as quasi-linear and composed of two parts: (i) a stand-alone valuation function that captures diminishing marginal utility within each individual product, and (ii) a bundle valuation that captures complementarities across products. To support estimation, we smooth the bidding data to ensure that each bidder’s demand sequence is monotonic, which is consistent with myopic bidding behavior. This monotonicity assumption is particularly helpful for identifying diminishing marginal utilities, as it removes non-structural “bidding in and out” fluctuations that arise from strategic adjustments. These modeling choices necessarily abstract from the full complexity of bidder behavior; however, the goal is not to exactly replicate the auction outcomes, but rather to construct a model that is sufficiently predictive to support meaningful counterfactual comparisons under alternative auction designs.

Under these assumptions, bidder valuations are estimated through a linear programming formulation that enforces several behavioral consistency constraints: nonnegative utilities, diminishing marginal returns, and revealed-preference relationships across rounds. The solution yields lowest valuation estimates compatible with the observed data, providing a coherent, lower-bound approximation of each bidder’s private values under myopic, monotonic behavior.

Once valuations are estimated, we simulate the auction to replicate observed bidding dynamics and validate the inferred valuations. Each bidder follows a \emph{myopic bidding strategy}, selecting in every round the bundle that maximizes their current utility given the prevailing price vector and their estimated valuations. We then test the model’s validity by comparing the results of simulated and actual auction, finding that it closely reproduces observed allocations while underestimating total revenue by only 6--10\%. The resulting outcomes confirm that the estimated valuations capture the fundamental structure of bidder behavior and provide a credible foundation for subsequent counterfactual policy analysis.

Having established a reliable estimation of bidder valuations, we perform a counterfactual analysis to address a specific policy question: how would embedding deployment commitments directly into the auction design affect revenue and coverage? In Canada’s current framework, deployment obligations are imposed ex post through population thresholds determined via consultation, often shaped by incumbent influence. To remedy this, we propose an extended auction mechanism where bidders select among low, medium, and high deployment levels within the auction itself. By linking discounted license prices to stronger deployment commitments, this design elicits truthful information about deployment costs and aligns private incentives with public coverage objectives.

Using our estimated valuations for five major operators, we simulate this extended mechanism and find that incorporating deployment obligations within the auction increases high-deployment license allocations by roughly 20\%, extending broadband access to an additional 3.3~million Canadians. Although revenues fall modestly, yet less than the estimated additional deployment costs, the design achieves better alignment with policy goals, showing that targeted rule modifications can balance efficiency, equity, and fiscal trade-offs.

\paragraph{Contributions.}
This paper contributes in two principal ways. First, on the modeling side, it introduces a simple yet behaviorally grounded method to replicate complex national auctions with minimal assumptions, enabling efficient and interpretable counterfactual analysis. Second, on the policy side, it applies this method to demonstrate how embedding deployment incentives directly into the auction mechanism can deliver meaningful coverage improvements while also increasing transparency in how deployment requirements are determined. Together, these contributions advance both the methodological and practical foundations of data-driven auction design and evaluation.

\section{Canada’s 2023 Spectrum Auction}
In 2023, Innovation, Science, and Economic Development Canada (ISED) held a major spectrum auction for licenses in the 3800 MHz band, covering frequencies from 3650 to 3900 MHz \cite{ISED3800}.  The auction ran from October 24 to November 24, spanning 98 bidding rounds for 21 business days. A total of 4,099 licenses were awarded to 20 bidders, including national carriers and regional telecom providers. The auction raised approximately 2.16 billion Canadian dollars in revenue.

The 3800~MHz auction used a clock auction format that is commonly employed for spectrum allocation. More specifically, it followed a two-stage structure: a \emph{clock stage} followed by an \emph{assignment stage}. During the clock stage, the auctioneer offered generic spectrum blocks\footnote{Generic licenses are blocks of spectrum that are sufficiently similar and comparable in value to one another so that they can be offered as a single category in each service area.} in each geographic service area (called Tier~4 service areas), often referred to as products. These blocks are identical in terms of bandwidth and conditions of use but are not yet tied to specific frequency positions. 

At the start of each round, the auctioneer announces a start price and a clock price for every product. The start price is the minimum price at which a product can be offered in that round, and in the first round it is set by ISED based on reserve prices. The clock price is calculated by applying a percentage increment, typically around 10\%, to the start price. 

Within each round, bidders express their demand by indicating, for each product, the number of generic blocks they wish to acquire at prices between the start and clock price (including both end points). After the bidders submit their demands, the auctioneer determines the posted price for each product based on aggregate demand. If the aggregate demand for a product is greater than its supply, the posted price is set to the clock price; otherwise, the posted price remains at the start price. The posted price for each product becomes the start price in the next round. In this way, prices increase only when there is excess demand. Bidders respond to the auctioneer by submitting the number of blocks they wish to purchase at announced prices. To prevent bidders from gaming the system by withholding activity early on, the auction uses activity rules: bidders must remain active across rounds or lose their eligibility to bid on large quantities in future rounds. Importantly, the clock stage is intra-round anonymous, meaning bidders do not see what others are bidding, only the aggregate demand after each round.

Once the clock stage concludes and winners are determined for each product (i.e., bidders who secured a certain quantity of generic blocks in specific service areas), the auction moves to the assignment stage. This stage determines the specific frequencies assigned to each winner among the blocks they won. The auction ends when, for all products in the auction, the aggregate demand is less than or equal to the available supply. At this point, no further rounds are held, and the final posted prices are locked in for each product. The number of blocks allocated to each bidder is determined based on their final demands at the posted prices. Having described the structure of the 3800 MHz auction, we now turn to the modeling process. 

\section{Preliminaries}
\label{sec:preliminaries}
Before presenting our analysis, we introduce the key components of the auction environment and the notation used throughout the paper. This section formalizes the structure of the clock auction and the representation of bids. The assumptions underlying bidder strategies will be introduced in Section~\ref{sec: model}.

A \textit{clock auction} is a multi-round process involving a set of bidders $I = \{1, \dots ,n\}$ and a set of products $J = \{1, \dots, m\}$, where each product represents a spectrum block in a specific geographic region. The auction proceeds over a finite set of rounds $\mathcal{R} = \{1, \dots ,R\}$. At each round $r \in \mathcal{R}$, the auctioneer announces a \textit{start price} $q_j^r$ and a \textit{clock price} $w_j^r$ for each product $j \in J$. The start price is the minimum price at which product $j$ can be offered in round $r$, and the clock price is computed as
\begin{eqnarray}
	w_j^r = q_j^r (1 + \delta_j^r),  
\end{eqnarray}
where $\delta_j^r \in [0.1, 0.2]$ is the price increment percentage for product $j$ at round $r$. The clock price represents the maximum price that product $j$ may reach in round $r$.

After the auctioneer announces $q_j^r$ and $w_j^r$, bidders simultaneously submit their bids. A bid vector for bidder $i$ at round $r$ is denoted by $b_i^r = (c_{i,1}^r, \dots, c_{i,m}^r)$, where $c_{i,j}^r \in \{0, \dots, s_j \}$ indicates the number of units bidder $i$ demands of product $j$, and $s_j$ is the available supply of product $j$. The \textit{aggregate demand} for product $j$ at round $r$ is given by
\begin{eqnarray}
	a_j^r = \sum_{i \in I} c_{i,j}^r,
\end{eqnarray}
and the aggregate demand vector is denoted $a^r = (a_1^r, \dots, a_m^r)$. If $a_j^r \leq s_j$, product $j$ is said to have cleared the market and its price remains at the start price $q_j^r$. Otherwise, the product is considered overdemanded and its price is raised to the clock price $w_j^r$. The final price for each product at round $r$ is called the \textit{posted price} and is denoted by $p_j^r$. The vectors of start, clock, and posted prices at round $r$ are denoted by $q^r$, $w^r$, and $p^r$, respectively.

The payment that bidder $i$ must make at the end of the auction (round $R$) is determined by the prices and quantities in their final allocation:
\begin{eqnarray}
	\sum_{j \in J} p_j^R \cdot b_{i,j}^R.
\end{eqnarray}
The auction terminates once all products clear simultaneously, i.e., when $a_j^r \leq s_j$ for every $j \in J$.

Each bidder $i$ is assumed to select in each round $r$ a bundle from a finite set $\mathcal{B}_i$ of feasible combinations of product units. Let $|\mathcal{B}_i| = L_i$, with $L_i \leq \prod_{j \in J} s_j$. Each bidder enters the auction with a valuation function $v_i: \mathcal{B}_i \rightarrow \mathbb{R}^+$, which assigns a non-negative value to every bundle. If bidder $i$ wins bundle $b_i^R$ at the end of the auction, their final utility is defined as
\begin{eqnarray}
	u_i(b_i^R) = v_i(b_i^R) - \sum_{j \in J} p_j^R \cdot b_{i,j}^R.
\end{eqnarray}

A \textit{strategy} for bidder $i$ in a clock auction is a function $\sigma_i: \mathcal{H}^r \rightarrow \mathcal{B}_i$, where $\mathcal{H}^r$ denotes the history available to the bidder at round $r$. This includes all posted price vectors $\{p^1, \dots, p^{r-1}\}$, aggregate demand vectors $\{a^1, \dots, a^{r-1}\}$, and the bidder’s own past bids $\{b_i^1, \dots, b_i^{r-1}\}$, along with the current round's start and clock prices $q^r$ and $w^r$.

Even if we restrict $\sigma_i$ to depend only on $q^r$ and $w^r$, and further assume a constant price increment $\bar\delta_j$ for each product across all rounds, the size of the strategy space is already exponential in $r$. If a bidder can choose from up to $\prod_{j \in J} s_j$ bundles per round, the number of possible strategies would be
\begin{eqnarray}
	\left( \prod_{j \in J} s_j \right)^{r^m}.
\end{eqnarray}
Hence, the full strategy space
\begin{eqnarray}
	\Sigma_i = \{ \sigma_i: \mathcal{H}^r \rightarrow \mathcal{B}_i \}
\end{eqnarray}
is vastly larger, satisfying
\begin{eqnarray}
	\left( \prod_{j \in J} s_j \right)^{r^m} \ll |\Sigma_i|.
\end{eqnarray}

\noindent In summary, the strategy space in a clock auction grows at an astronomical rate with the number of products, rounds, and feasible bundles. This combinatorial explosion makes it infeasible to characterize or simulate bidding behavior directly from first principles. For this reason, any empirical or counterfactual study of spectrum auctions must rely on simplifying assumptions and structured models of bidder behavior to yield tractable and interpretable results.

\section{Recreating the 3800 MHz Auction}
\label{sec: model}
To replicate the behavior observed in Canada’s 3800 MHz auction, this section outlines the modeling and simulation framework developed for our study. We begin by describing the key behavioral and structural assumptions that make the estimation problem tractable while preserving the essential dynamics of the real auction. Based on these assumptions, we formulate a linear programming model that infers bidders’ latent valuations from their observed round-by-round bidding behavior. These estimated valuations then serve as inputs to our simulation framework, which reconstructs the auction process and enables counterfactual analysis under alternative policy scenarios.
\subsection{Valuation Estimation}
\label{sec: valuation_estimation}
To simulate Canada’s 3800 MHz auction and assess alternative policy designs, we first require a working model of bidders’ valuations. These valuations are not directly observable but can be inferred from the round-by-round bidding data available for the auction.\footnote{All the data used in this study is available at ISED's website\cite{ISED3800}.}

Spectrum auctions—particularly formats like the Simultaneous Multiple Round Auction (SMRA) and Combinatorial Clock Auction (CCA)—pose substantial modeling challenges due to their dynamic nature, multi-item structure, and strategic complexity. To make simulation and counterfactual analysis tractable, we impose a set of simplifying assumptions that preserve key features of the original environment while reducing its strategic and computational complexity.

\paragraph{Myopic Bidders.}  
We assume bidders are myopic: in each round, they select the bundle that maximizes their immediate utility given the current prices, without considering how their bids may affect future prices, rounds. Under this assumption, the bundle observed in round $r$ at a price vector $p^r$ must yield weakly higher utility than the bundles chosen in rounds $r{-}1$ and $r{+}1$, when evaluated at the same price vector $p^r$. 

\paragraph{Restricted Bundle Set.}  
Rather than allowing bidders to choose from the full combinatorial space of product bundles, we restrict each bidder’s feasible set to the bundles they actually submitted bids for during the real auction. This is both a practical simplification and a modeling assumption supported by prior findings that spectrum bidders often pre-select a small set of preferred packages before entering the auction \cite{Scheffel2012}. We define a \textit{bundle base} for a given bundle as the combination of products that appear in that bundle, where each product is assigned its minimum number of copies demanded by the bidder during the auction. In other words, the bundle base represents the smallest observed configuration of a bidder’s interest across the products included in the bundle. For example, consider a bundle that includes only products~A and~B. If a bidder demanded $(A{:}4, B{:}5)$ (that is, four copies of~A and five copies of~B) and their minimum demand during the auction was four for~A and three for~B, then the bundle base associated with this bundle would be $(A{:}4, B{:}3)$. Variants of these bases are generated by increasing the quantity of one or more products up to the maximum observed demand. Note that this assumption does not materially alter the auction outcome in any meaningful way, but it substantially reduces the size of the bundle space and makes the estimation problem feasible.

\paragraph{Valuation Structure.}  
We model each bidder’s valuation as quasi-linear, consisting of two components. The first captures how a bidder’s willingness to pay for additional copies of the same product gradually decreases, reflecting diminishing marginal returns. The second captures the added value that arises when certain products are acquired together, what we refer to as \emph{bundle complementarities}. 
Rather than assuming a fully additive or separable structure, we allow these complementarities to emerge directly from the data. For every bundle observed in the auction, we assign a distinct valuation parameter and infer its value based on the bidder’s observed choices and the prevailing prices. This approach provides a flexible representation of bidder preferences: it captures both within-product diminishing returns and cross-product synergies without requiring strong structural assumptions about how these effects interact.

\paragraph{Monotonic Bidding.}  
To align with the myopic bidding assumption, we smooth the bidding data to enforce monotonicity in demand: a bidder cannot reduce demand for a product and later increase it. Non-monotonic patterns, typically indicative of forward-looking or strategic behavior, are removed during preprocessing. This assumption follows naturally from the notion of straightforward bidding, as bidders who respond myopically to prices are expected to reduce demand only once and not revert to higher levels later. Moreover, maintaining monotonicity is essential for the tractability of our estimation procedure. The diminishing-returns constraints used to infer stand-alone valuations rely on the idea that once a bidder ceases to request an additional copy of a product, its marginal value must be below the price. Allowing reversals in demand would violate this logic and render the resulting linear program infeasible. 

\paragraph{Estimation via Linear Programming.}
Given these assumptions, we estimate valuations by solving a linear program. The LP for each bidder $i$ is independent, so we drop the subscript $i$. 

For each product $j$, let 
\begin{eqnarray}
	C_j = \{ c_{j,1}, c_{j,2}, \dots, c_{j,K_j} \}, 
\end{eqnarray}
be the ordered set of distinct quantities demanded during the auction (monotonic by construction). We define $v_j^{(c_{j,k})}$ as the marginal valuation of increasing demand from $c_{j,k-1}$ to $c_{j,k}$. We assume diminishing returns:
\begin{eqnarray}
	v_j^{(c_{j,k})} \geq v_j^{(c_{j,k+1})}, \quad \forall k, k\neq 1.
\end{eqnarray}
For a bundle $b^r = (c_{1,k_1^r},\dots,c_{m, k_m^r})$ played at round $r$ with base $\bar b$, we define
\begin{eqnarray}
	v(b^r) = v_{\bar b} + \sum_{j \in J_{b^r}} \sum_{l \in K_j, l\leq k_j^r} (c_{j,l}-c_{j,l-1}) \cdot v_j^{(c_{j,l})},
\end{eqnarray}
where $J_{b^r}$ denotes the set of products present in bundle $b^r$, and $v_{\bar b}$ captures complementarities across the products in our base. The variables $v_{\bar b}$ and $v_j^{(c_{j,l})}$ are the decision variables of the LP. Note that $v_j^{(c_{j,1})}$ corresponds to the value of holding the minimum number of copies of product~$j$ observed in the auction. This value reflects the baseline stand-alone component rather than the diminishing return from additional copies. As we are primarily interested in relative valuations, we normalize this baseline to zero.
\\
Then the utility of a bundle $b^r$ is calculated by 
\begin{eqnarray}
	u(b^r;p^r) = v_{\bar b} + \sum_{j \in J_{b^r}} \sum_{l \in K_j, l\leq k_j^r} (c_{j,l}-c_{j,l-1}) \cdot v_j^{(c_{j,l})} - c_{j,k_j^r}\cdot p_j^r,
\end{eqnarray}
where $p_j^r$ represents the start price of item $j$ at round $r$.
\\ 
We then solve the following LP:
\begin{align} 
	\min \quad & \sum_{b,b',r} S_{bb'r} + \sum_{\bar b} v_{\bar b} \\
	\text{s.t.} \quad 
	& \text{Positive utility:}\\
	& u(b^r; p^r) \geq 0 \quad \forall r \in \mathcal{R} \quad \\
	& \text{Marginal rationality:}\\
	&\sum_{l \in K_j, l\leq k_j^r} (c_{j,l} - c_{j,l-1}) v_j^{(c_{j,l})} - c_{j,k_j^r} \cdot p_j^r \geq \sum_{l \in K_j, l\leq k'_j}(c_{j,l} - c_{j,l-1}) v_j^{(c_{j,l})} - c_{j,k'_j} \cdot p_j^r \quad \\
	&\quad \forall r \in \mathcal{R} \quad \forall j\in J_{b^r}, \enspace \forall c_{j,k'_j} \in \{ \max{\{c_j \in C_j | c_j < c_{j,k_j^r}\}}, \min{\{c_j \in C_j | c_j > c_{j,k_j^r} \}} \} \quad  \\\nonumber 
	&\text{Revealed preferences:}\\
	&u(b^r;p^r) \geq u(b';p^r) - S_{bb'r} \quad \\
	&\forall r \in \mathcal{R}, \enspace  \forall b' \in \mathcal{B} \enspace \text{s.t.} \enspace b'\neq b^r, \enspace  El(b') \leq E^r \quad\\ \nonumber 
	& \text{Bounds:}\\
	& v_{\bar b} \geq 0 \quad \forall \bar b \in \mathcal{\bar B}\\
	& v_{j}^{(c_{j,k})} \geq 0 \quad \forall j \in J_{b^r}, \enspace \forall k \in K_j\\
	& S_{bb'r} \geq 0 \quad \forall b \in \mathcal{B}, \enspace \forall r \in \mathcal{R}, \enspace   \forall b' \in \mathcal{B} \enspace \text{s.t.} \enspace b'\neq b^r,\enspace El(b') \leq E^r
\end{align}

Here, the positive utility constraints ensure that every bundle observed during the auction must yield non-negative utility. The marginal rationality shows that if the bidder chose $c_{j,k_j}$ copies of a product $j$ it must be that it offered higher utility than choosing any of the neighboring lower or higher number of copies. These constraints are important for capturing the diminishing returns effect. Moreover, the revealed preferences constraints require that any bundle $b$ observed at round $r$ must have had at least as much utility as alternative bundles $b'$ from earlier or later rounds, provided those bundles were feasible under the bidder’s eligibility. Therefore eligibility rules are incorporated by restricting the comparison of the utility of bundle~$b^r$ to alternative bundles~$b'$ that satisfy $\text{El}(b') \leq E^r$, where $\text{El}(b')$ denotes the eligibility cost of~$b'$ and $E^r$ represents the bidder’s eligibility in round~$r$. The slack variables $S_{bb'r}$ penalize inconsistencies in observed bidding and are minimized in the objective. Minimizing $\sum_{\bar b} v_{\bar b}$ prevents the model from arbitrarily inflating complementarity terms. The resulting solution therefore provides the tightest lower-bound valuations consistent with myopic, monotonic bidding.

\subsection{Simulation Algorithm}
\label{sec: simulation_algorithm}
Given the estimated valuations, we now formalize the simulation for each company's bidding behavior. Assuming bidders follow a straightforward (myopic) strategy, we define a bidder's strategy $\sigma$ as a function:
\[
\sigma(\bar V, V, \mathcal{\bar B}, p^r) \rightarrow \mathcal{B},
\]
where $\bar V$ denotes the set of estimated bundle valuations, $V$ is the set of marginal product valuations, $\mathcal{\bar B}$ is the set of bundle bases, and $p^r$ is the price vector at round $r$. The strategy selects a utility-maximizing bundle $b^* \in \mathcal{B}$ given current prices and valuation information. The algorithm proceeds as follows:

\begin{algorithm}[H]
	\SetAlgoNoLine
	\KwIn{Bundle valuations $\bar V$, marginal valuations $V$, bundle bases $\mathcal{\bar B}$, price vector $p^r$, eligibility $E^r$}
	\KwOut{Utility-maximizing bundle ${b^*}^r$}
	$u^* \leftarrow -\infty$; \quad ${b^*}^r \leftarrow \text{None}$\;
	\For{each $\bar b \in \mathcal{\bar B}$}{
		$b^* \leftarrow \texttt{BEST\_COPIES}(\bar b, p^r, E^r)$\;
		\If{$b^* \neq \text{None}$ \textbf{and} $u(b^*; p^r) > u^*$}{
			${b^*}^r \leftarrow b^*$\;
			$u^* \leftarrow u(b^*; p^r)$\;
		}
	}
	\If{$u^* \geq 0$}{
		\Return ${b^*}^r$\;
	}
	\Else{
		\Return None\;
	}
	\caption{Myopic Bidding Strategy}
	\label{alg:one}
\end{algorithm}

The core subroutine \texttt{BEST\_COPIES} identifies the utility-maximizing assignment of copy counts within a given bundle base $\bar b$ using a Mixed Integer Program (MIP). Let $I_{j,k} \in \{0, 1\}$ be a binary decision variable indicating whether copy level $c_{j,k} \in C_j$ is chosen for product $j$. The optimization problem is formulated as:

\begin{align}
	\max \quad & \sum_{j} \sum_{k=2}^{K_j} I_{j,k} \cdot \left[ (c_{j,k} - c_{j,k-1}) \cdot v_j^{(c_{j,k})} - c_{j,k} \cdot p_j^r \right] \\
	\text{s.t.} \quad 
	& \text{Eligibility Constraint:}\\
	& \sum_{j} \sum_{k=2}^{K_j} I_{j,k} \cdot c_{j,k} \cdot e_j \leq E^r \quad  \\
	& \text{Copy limitation:}\\
	& \sum_{k=2}^{K_j} I_{j,k} = 1 \quad \forall j \in \bar b \quad \\
	& \text{Bound:}\\
	& I_{j,k} \in \{0,1\} \quad \forall j,k
\end{align}

Here, $v_j^{(c_{j,k})}$ is the marginal valuation of product $j$ at copy count $c_{j,k}$, $p_j^r$ is the price of product $j$ in round $r$, $e_j$ is the eligibility weight of product $j$, provided by the auctioneer, and $E^r$ is the bidder’s eligibility score at round $r$.

The objective seeks to maximize the total utility by choosing one copy count level per product such that the total eligibility cost does not exceed $E^r$. Note that bundle valuations $\bar V$ are used only in the outer strategy function (Algorithm~\ref{alg:one}) to compare utility across bundles, while the MIP operates solely on marginal valuations $V$.

\section{Model Evaluation}

To evaluate our model, we simulate the 3800 MHz clock auction using rules that closely replicate the actual auction format, with bidders playing the strategies described in Section~\ref{sec: simulation_algorithm} and valuations estimated using the method in Section~\ref{sec: valuation_estimation}. We refer to the actual auction as \textit{Auction A} and the simulated auction as \textit{Auction B}. As an additional robustness check, we apply the same simulation procedure to the 3500 MHz auction by the end of this section.

The primary goal of this exercise is to assess how well our model replicates the outcomes of Auction $A$. Naturally, this process comes close to overfitting, since the same dataset used to calibrate the model is also used for validation. However, this limitation is not problematic for our purposes. Our goal is not generalization to other auctions, but rather counterfactual analysis of Auction $A$ itself. Internal validity, accurately capturing the incentives, bidding dynamics, and price formation mechanisms in this specific setting is what matters most.

That said, validation remains crucial. To verify that the model captures the core behavioral and structural patterns observed in Auction A, we compare outcomes from Auctions $A$ and B across several dimensions: final allocations, total revenue, and bidding trajectories. Consistency across these metrics provides evidence that the model offers a faithful computational representation of the real auction and is suitable for counterfactual experiments.

\begin{figure}[htbp]
	\centering
	\begin{subfigure}[b]{0.49\textwidth}
		\includegraphics[width=\linewidth]{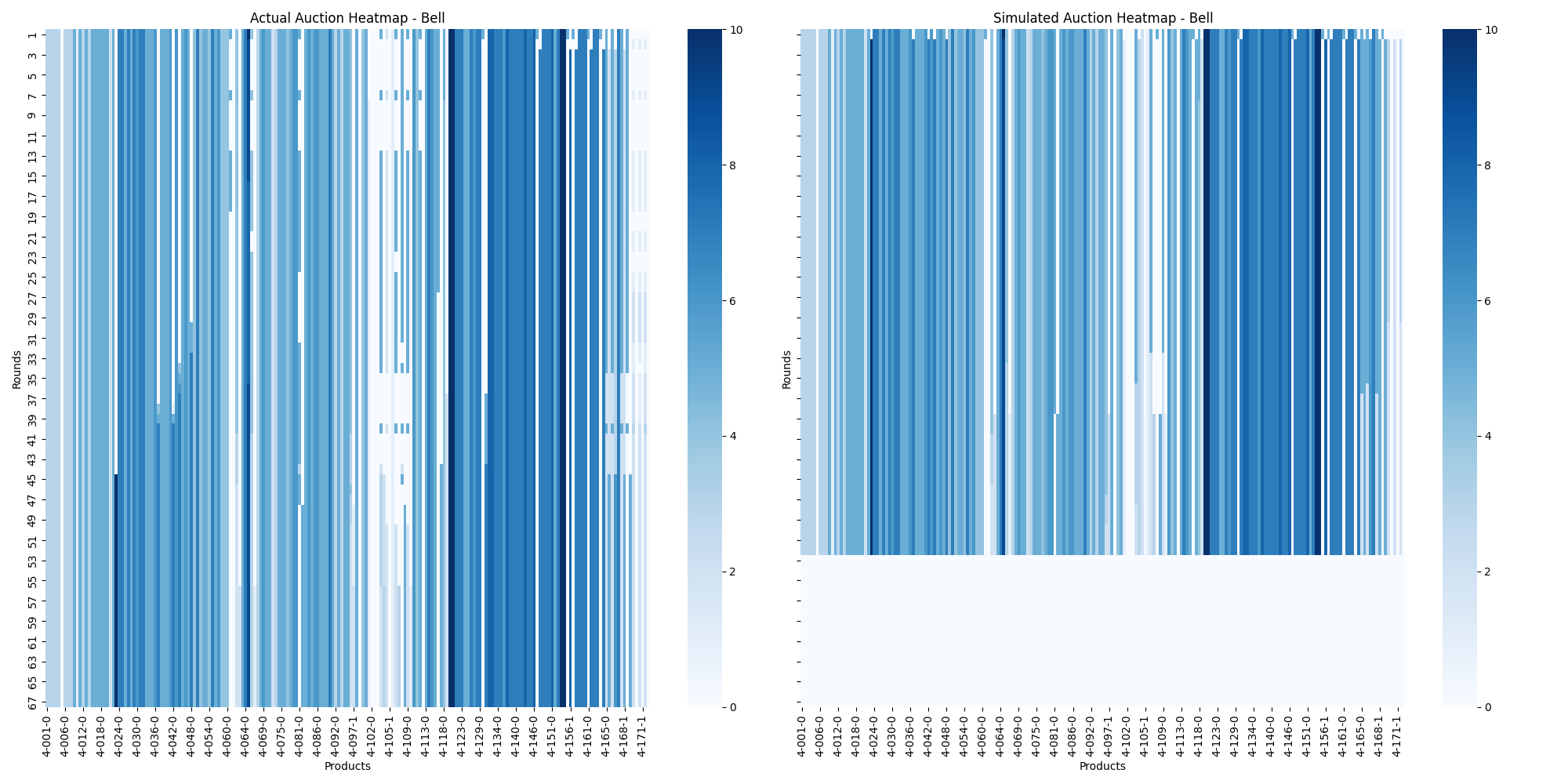}
		\caption{Bell}
		\label{fig:Bell_heatmap}
	\end{subfigure}
	\begin{subfigure}[b]{0.49\textwidth}
		\includegraphics[width=\linewidth]{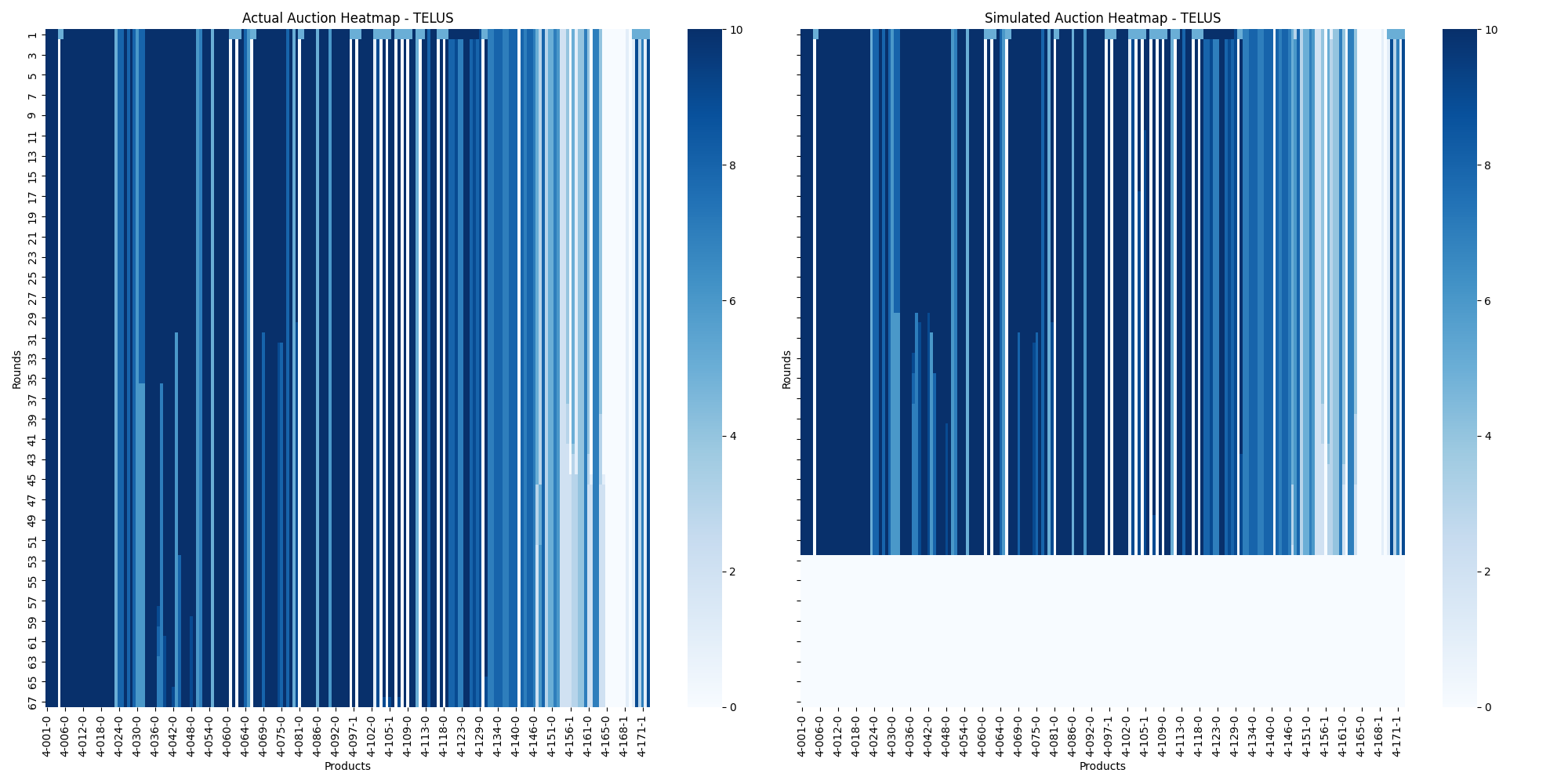}
		\caption{TELUS}
		\label{fig:TELUS_heatmap}
	\end{subfigure}
	\begin{subfigure}[b]{0.49\textwidth}
		\includegraphics[width=\linewidth]{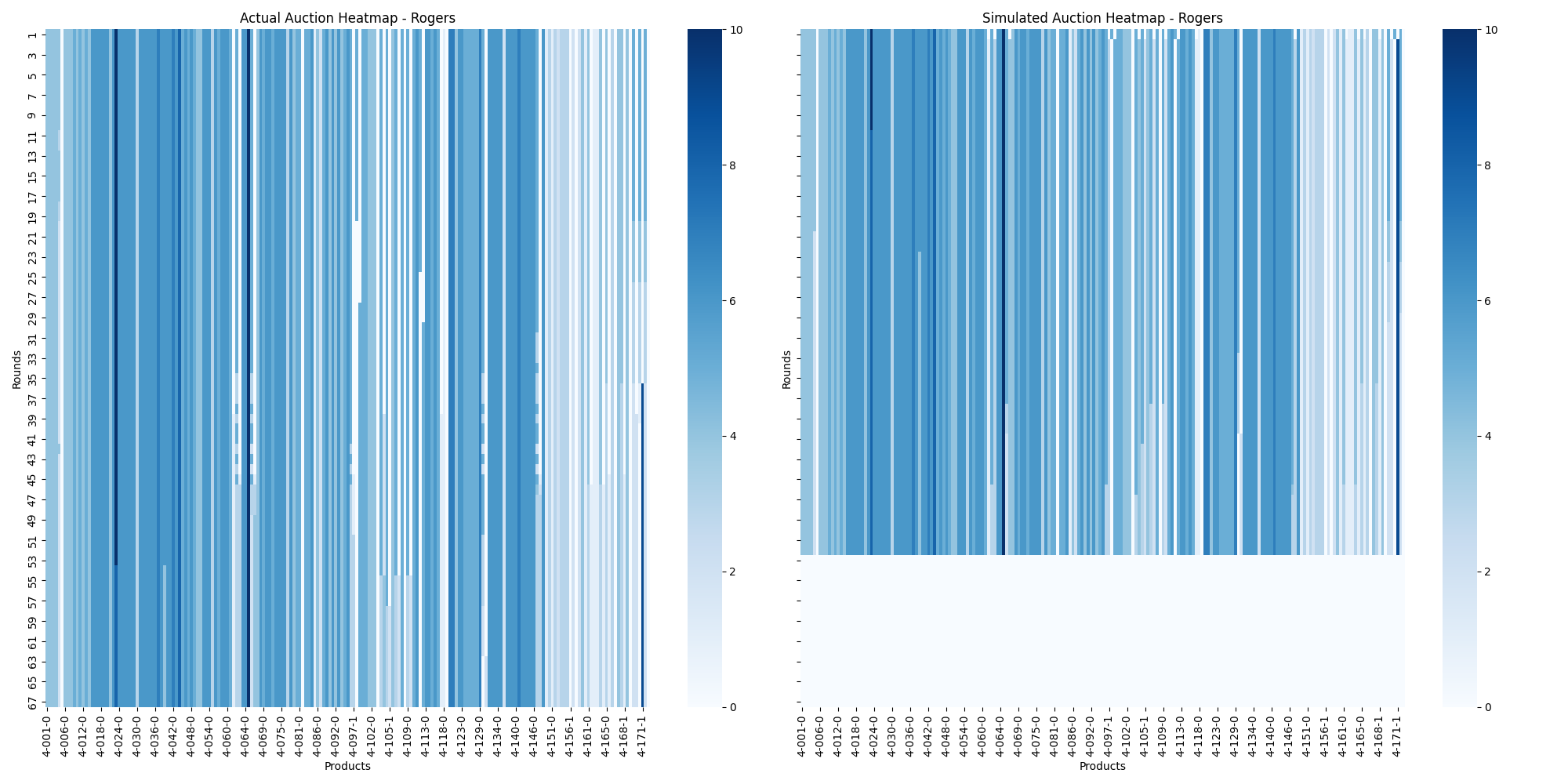}
		\caption{Rogers}
		\label{fig:Rogers_heatmap}
	\end{subfigure}
	\begin{subfigure}[b]{0.49\textwidth}
		\includegraphics[width=\linewidth]{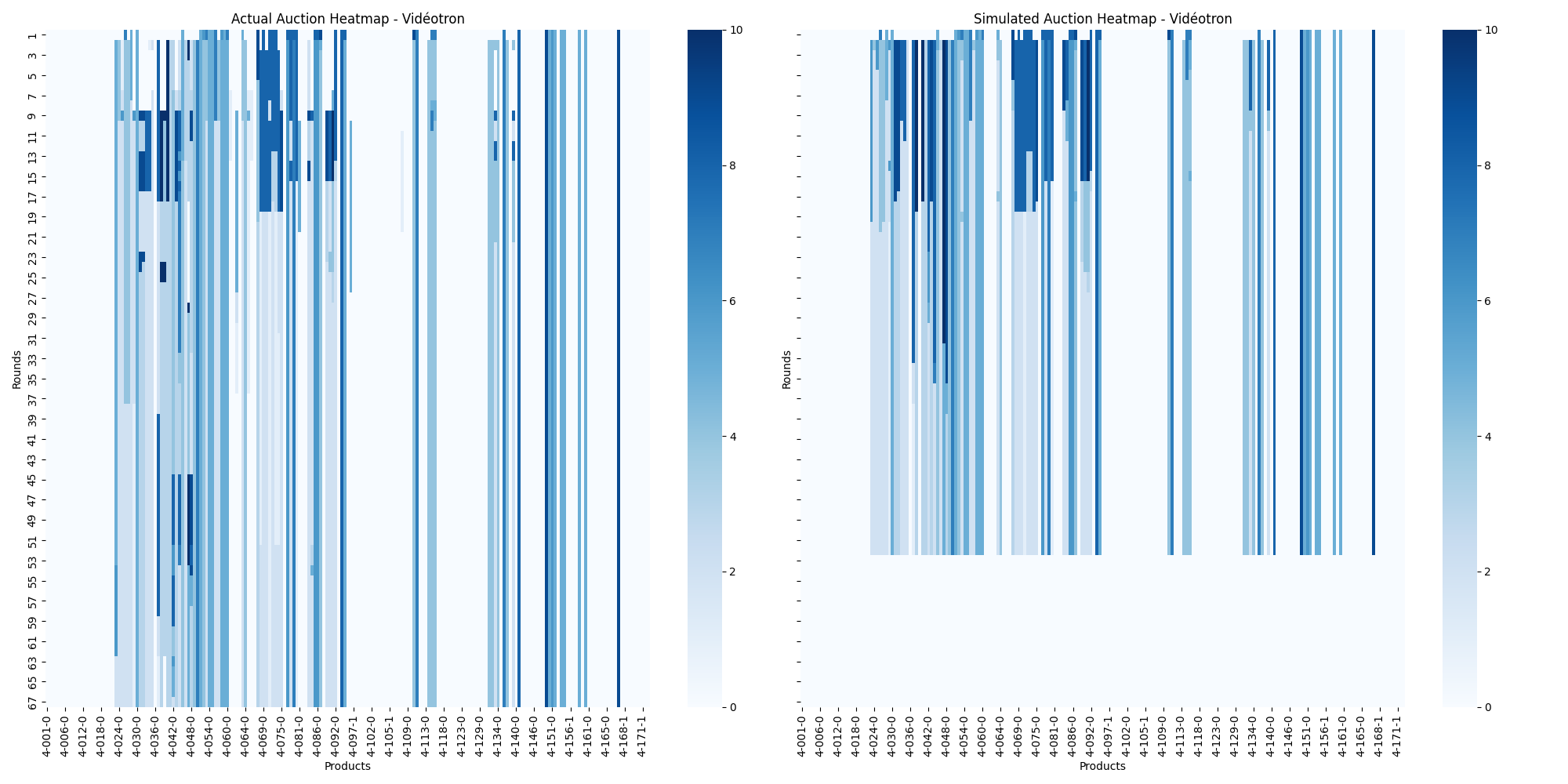}
		\caption{Vidéotron}
		\label{fig:Vidéotron_heatmap}
	\end{subfigure}
	\begin{subfigure}[b]{0.49\textwidth}
		\includegraphics[width=\linewidth]{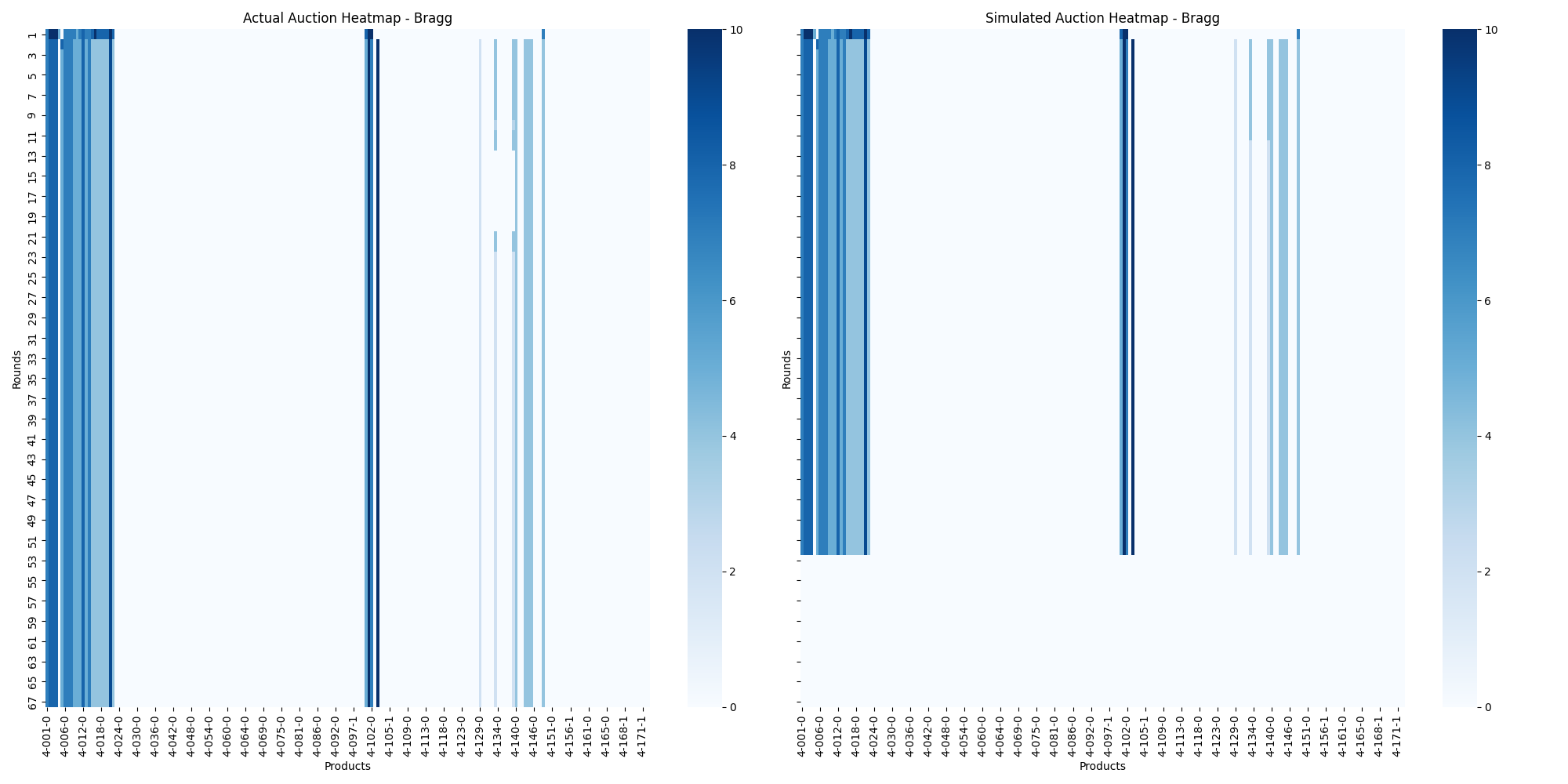}
		\caption{Bragg}
		\label{fig:Bragg_heatmap}
	\end{subfigure}
	\caption{Bidding patterns visualized as heatmaps. Left: Auction $A$ (actual); Right: Auction $B$ (simulated).}
	\label{fig:heatmaps}
\end{figure}

While we do not aim to replicate the auction round-by-round, the round-level comparisons still yield valuable insights into bidder behavior. Figures~\ref{fig:Bell_heatmap}--\ref{fig:Bragg_heatmap} show heatmaps of the bundles bid on in each round. The left side shows bids in Auction A, and the right side shows Auction B. We present these heatmaps for five companies, Bell, TELUS, Rogers, Vidéotron, and Bragg, which we include in our counterfactual analysis.

From these heatmaps, we observe that Auction $B$ ends earlier (in 52 rounds) than Auction $A$ (which concludes in 64 rounds). Furthermore, bidders in Auction B exhibit smoother, more consistent bundle transitions over time, while Auction $A$ contains abrupt changes in demand. These sharp fluctuations likely stem from strategic behaviors not captured by our modeling assumptions. Nonetheless, such discrepancies are acceptable as long as the key outcomes remain aligned.

To quantify the similarity between final allocations, we compute the RMSE (root mean square error) between the final bundles in Auctions $A$ and $B$. The average RMSE across the five companies is 0.08. Rogers, Bell, Vidéotron, and Bragg all match exactly (RMSE = 0), while TELUS differs slightly (RMSE = 0.26), indicating a deviation of less than one copy per product on average. In total, Auction $A$ clears 4,099 units, compared to 4,088 in Auction $B$.

\begin{figure}
	\centering
	\includegraphics[width=0.5\linewidth]{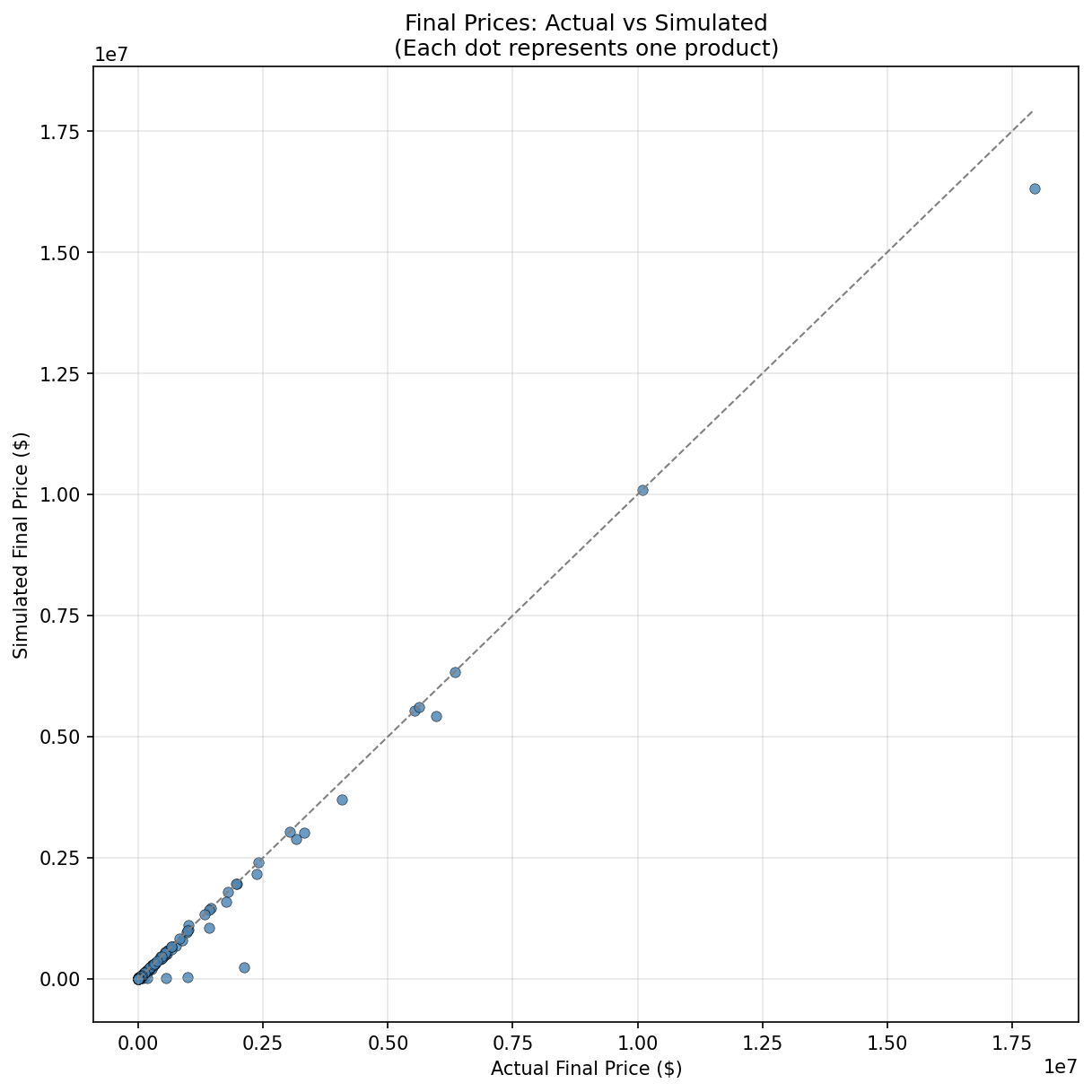}
	\caption{Final prices in Auction $B$ (simulated) relative to Auction $A$ (actual).}
	\label{fig:final_price_scatter}
\end{figure}

Another important dimension is the evolution of prices. In Figure~\ref{fig:final_price_scatter}, we present a scatter plot comparing the final prices of products in Auctions $A$ and $B$. While many products match their actual final price exactly, several reach lower final prices in the simulation. This is expected: our valuation estimation approach is designed to find the \textit{minimal} valuations consistent with observed bidding behavior. As a result, Auction $B$ clears at lower prices and concludes earlier.

Consequently, total revenue in Auction $B$ is \$1,903,834,500 compared to \$2,026,057,000 in Auction $A$, a 6\% shortfall. This gap highlights a key trade-off in our estimation: although the model reproduces allocations with high fidelity, it underestimates willingness to pay, due to its conservative (lower-bound) valuation approach.

Moreover, for the 3500 MHz auction that took place in 2021, there is also publicly available data from \cite{ISED3500}, which allows us to apply the same valuation estimation procedure and simulation algorithm used for the 3800 MHz auction. While this dataset cannot serve as an out-of-sample test set in a strict sense, given that the set of participating companies and their characteristics differ from those in the 3800 MHz auction, it nevertheless offers a valuable robustness check for our modeling framework. The primary participants in this auction were Bell, Rogers, TELUS, Vidéotron, and Xplornet.

\begin{figure}[htbp]
	\centering
	\begin{subfigure}[b]{0.49\textwidth}
		\includegraphics[width=\linewidth]{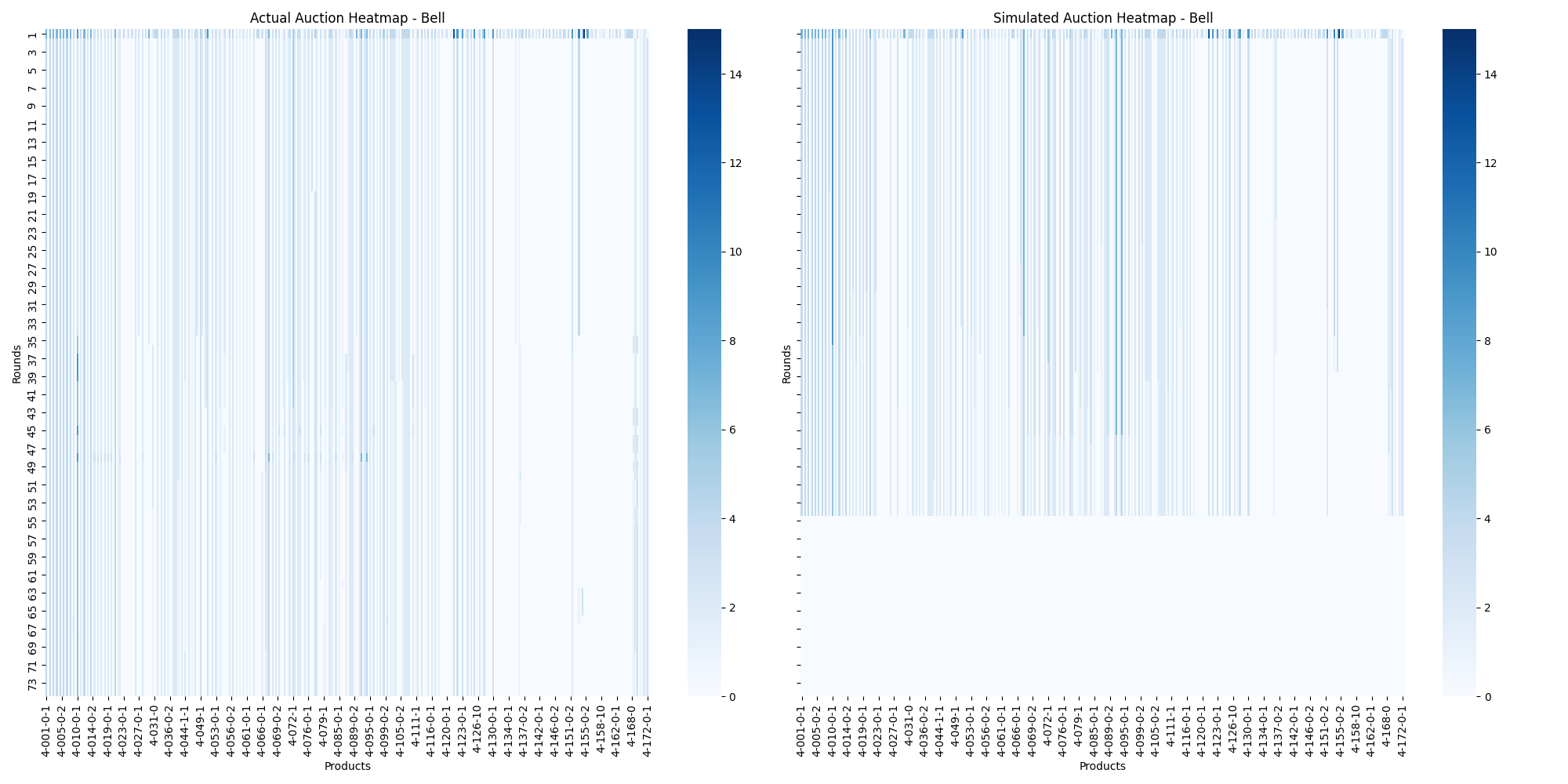}
		\caption{Bell}
		\label{fig:Bell_heatmap2}
	\end{subfigure}
	\begin{subfigure}[b]{0.49\textwidth}
		\includegraphics[width=\linewidth]{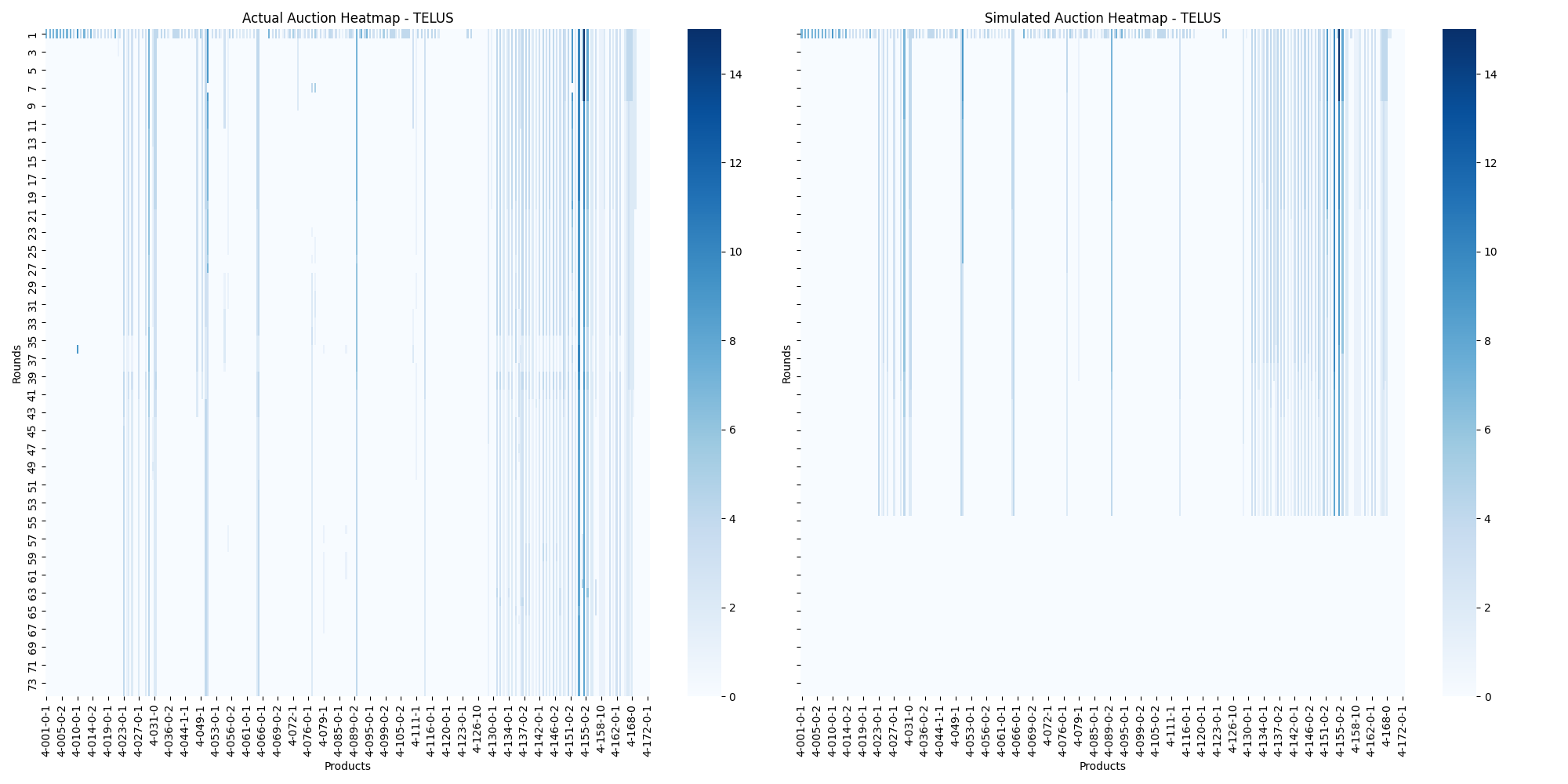}
		\caption{TELUS}
		\label{fig:TELUS_heatmap2}
	\end{subfigure}
	\begin{subfigure}[b]{0.49\textwidth}
		\includegraphics[width=\linewidth]{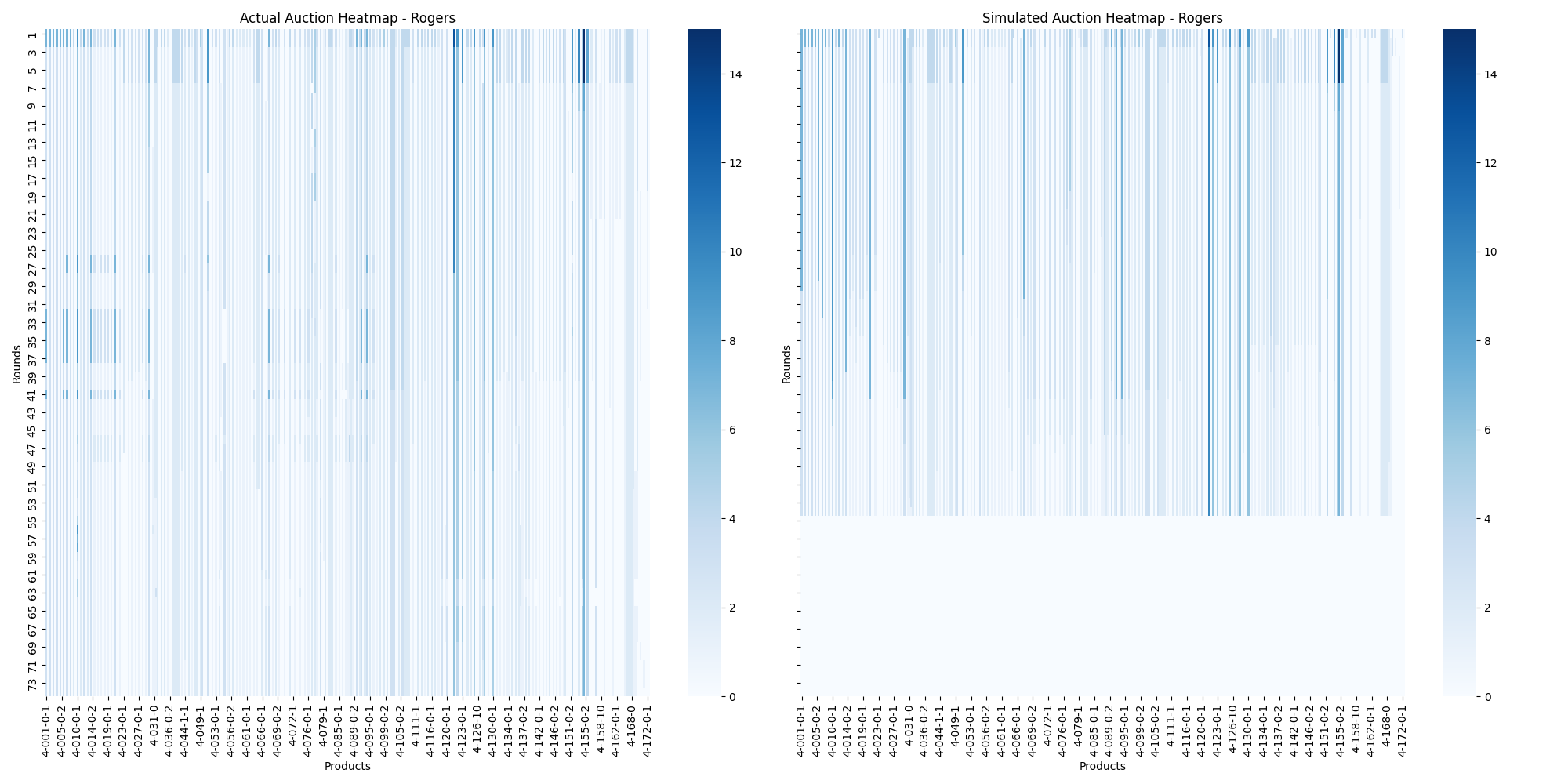}
		\caption{Rogers}
		\label{fig:Rogers_heatmap2}
	\end{subfigure}
	\begin{subfigure}[b]{0.49\textwidth}
		\includegraphics[width=\linewidth]{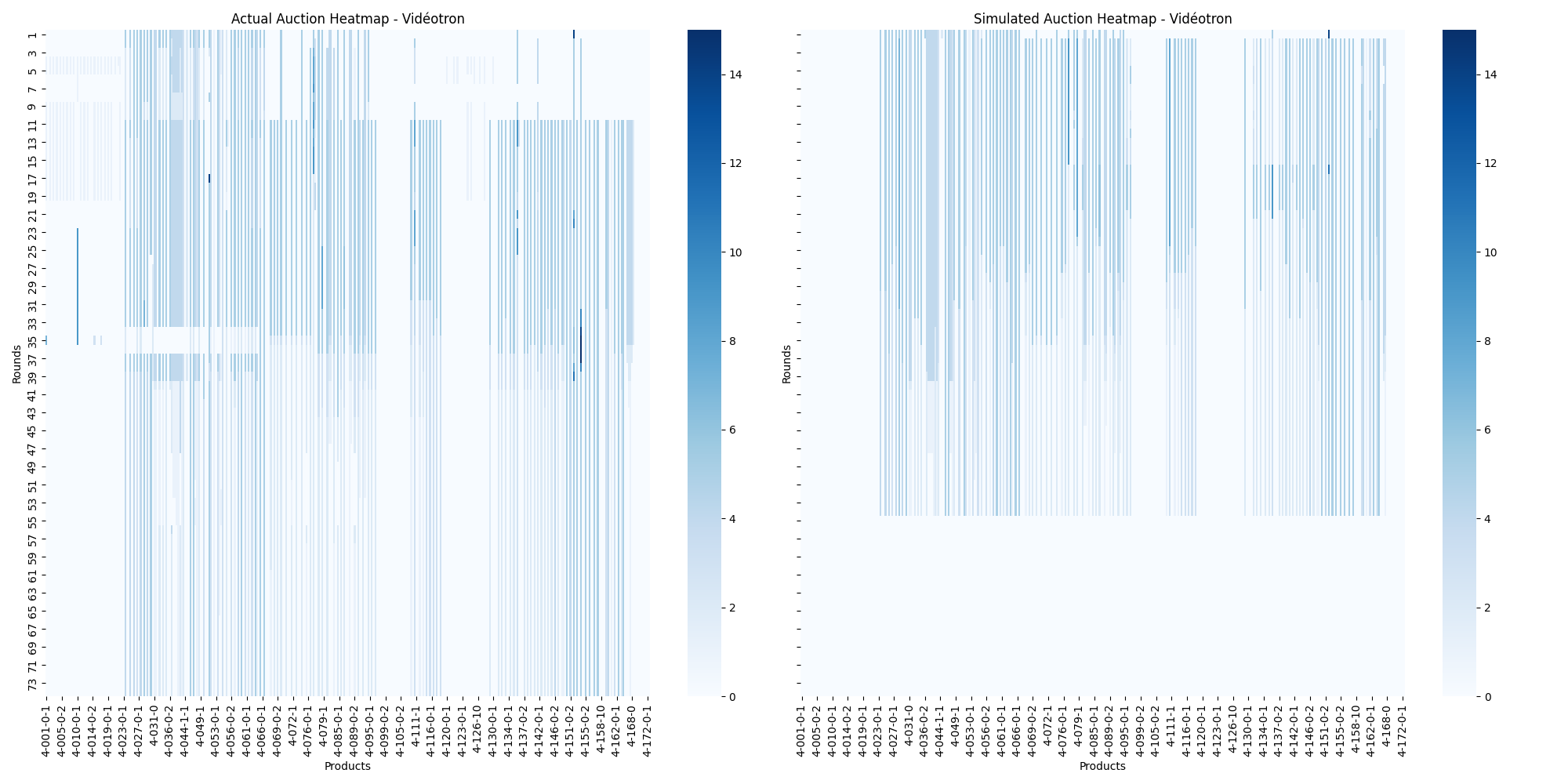}
		\caption{Vidéotron}
		\label{fig:Vidéotron_heatmap2}
	\end{subfigure}
	\begin{subfigure}[b]{0.49\textwidth}
		\includegraphics[width=\linewidth]{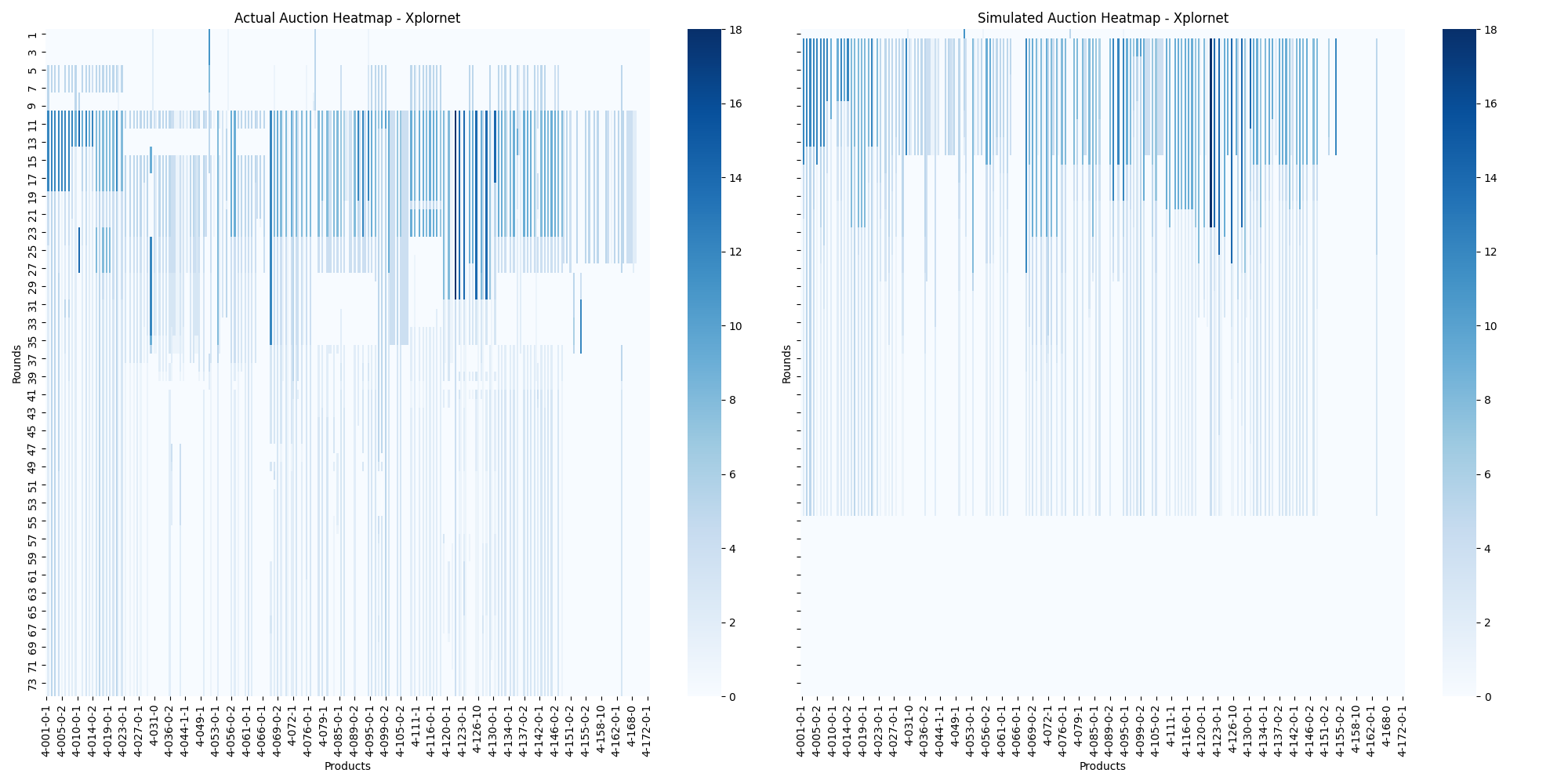}
		\caption{Xplornet}
		\label{fig:Xplornet_heatmap2}
	\end{subfigure}
	\caption{Bidding patterns visualized as heatmaps for the 3500 MHz auction. Left: Auction $C$ (actual); Right: Auction $D$ (simulated).}
	\label{fig:heatmaps}
\end{figure}

Figures \ref{fig:Bell_heatmap2}–\ref{fig:Xplornet_heatmap2} show the bidding patterns for these five bidders in the real auction $C$ (left) and the simulated auction $D$ (right). As in the 3800 MHz setting, we observe that bidding in the simulated auction is smoother and more monotonic, with fewer abrupt changes across rounds. The simulated auction also concludes earlier, at round 54 compared to round 74 in the actual auction. This pattern reflects the structure of our model, which relies on myopic, utility-maximizing bidding without behavioral noise or strategic experimentation.

Despite these simplifying assumptions, the model performs strongly in replicating final outcomes. The average RMSE between final bundles in auctions $C$ and $D$ is just 0.017 across all companies. For Bell, TELUS, Vidéotron, and Xplornet, final bundles match exactly, resulting in RMSEs of 0, while Rogers deviates slightly with an RMSE of 0.37. This small discrepancy accounts for the slight difference in total quantities sold: 14,066 in the actual auction versus 14,077 in the simulated auction.

\begin{figure}
	\centering
	\includegraphics[width=0.5\linewidth]{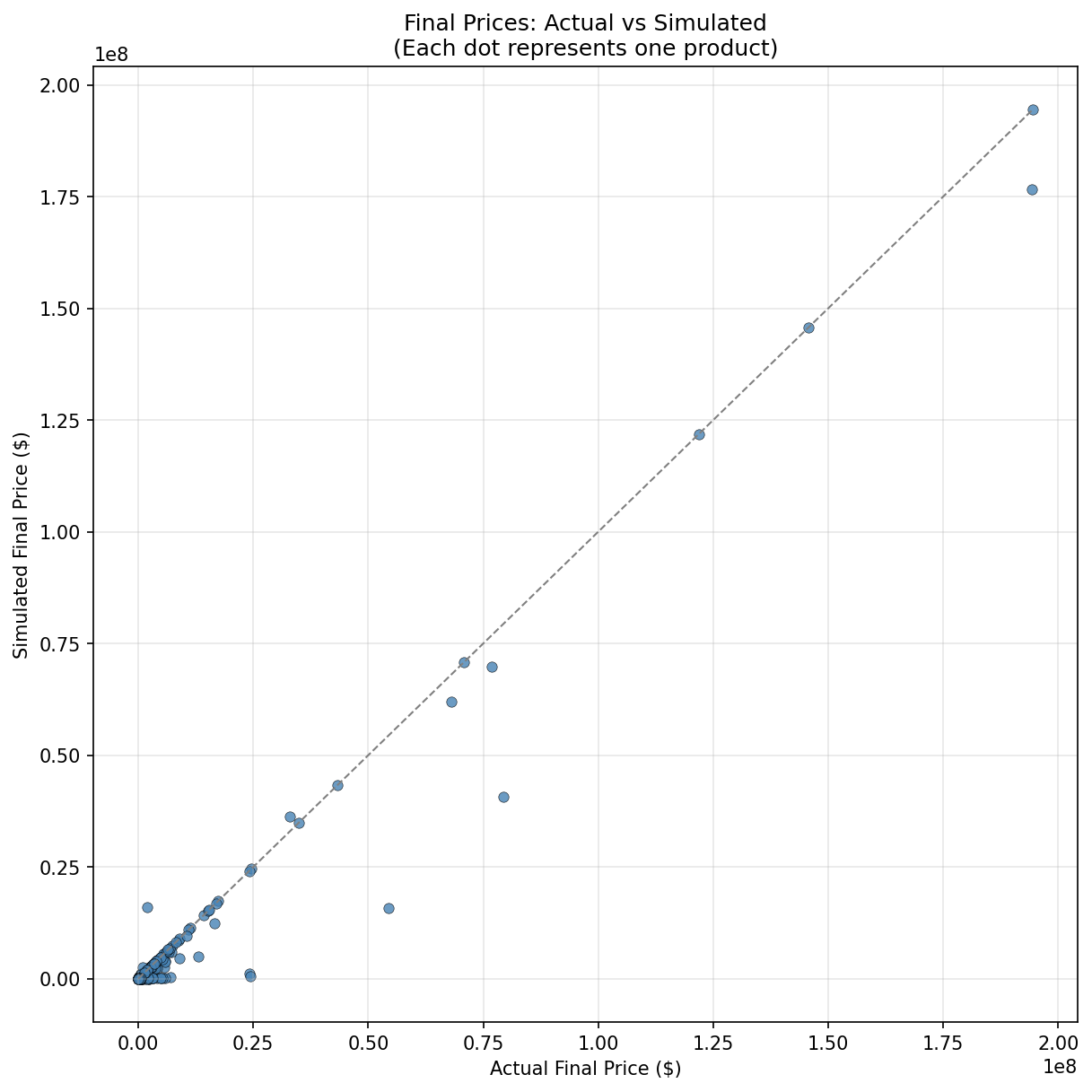}
	\caption{Final prices in Auction $D$ (simulated) relative to Auction $C$ (actual).}
	\label{fig:final_price_scatter2}
\end{figure}

Final prices in auction $D$ follow the same pattern as in the 3800 MHz simulation: while a large subset of products ends with prices close to the actual values, several are noticeably underpriced. As shown in Figure \ref{fig:final_price_scatter2}, most of the deviations fall below the 45-degree line. Consequently, the total revenue in the simulated auction is \$7,924,278,135 compared to \$8,831,029,798 in the real auction, a difference of roughly 10\%. This pattern again illustrates a feature of our estimation framework: by seeking minimal valuations consistent with observed behavior, the model tends to underpredict prices while still matching allocation patterns.

\section{Counterfactual Analysis of Deployment Requirements}

In our counterfactual analysis, we consider a modification to the standard clock auction format by embedding deployment requirements (also known as build-out or roll-out requirements in other jurisdictions) directly into the products being auctioned. Deployment requirements are set to ensure that the acquired spectrum is put to use by requiring licensees to provide service to a minimum percentage of the population within the licensed area (as little as 5\% of the population in service area up to 70\%) in a given timeframe.  Each product in the extended auction is associated with one of three deployment levels: \textit{Low}, \textit{Medium}, and \textit{High}, corresponding to ISED’s 5-\footnote{ISED applied two different term lengths for the shortest set of deployment requirements. In Tier 4 areas containing large population centres (24 areas in total), deployment obligations had to be met within five years, whereas in areas without large population centres (148 areas) they could be met over a slightly longer, seven-year period. For simplicity, we refer to both cases as the five-year requirements throughout.}, 10-, and 20-year deployment thresholds, respectively. These obligations are \textit{time-compressed}: bidders committing to a higher deployment level must fulfill longer-term coverage obligations within just five years. For example, selecting the \textit{High} variant requires completing, within five years, what would typically be required over a 20-year license duration.

In the extended auction, each deployment level has its own demand and price adjustment rule, while all levels share a common supply constraint for the underlying product. After each round, the aggregate demand at each deployment level is compared with the available supply to determine whether the product is overdemanded. Importantly, the deployment levels are hierarchically related: stricter (higher) levels imply all weaker ones. For example, a commitment to the high deployment level automatically satisfies the requirements of the medium and low levels. A deployment level is considered overdemanded if the aggregate demand for that level, or for any combination of levels that imply it, exceeds the available supply. Prices are then updated accordingly:

\begin{itemize}
	\item For high-level products, the posted price increases only if demand at the high level exceeds supply. Otherwise, the price remains unchanged.
	\item For medium-level products, the price increases if either the medium or high levels (or their combination) are overdemanded. 
	\item For low-level products, the price increases if demand exceeds supply in any level or combination of levels. 
\end{itemize}

This tiered price-adjustment structure creates a built-in incentive gradient: as higher deployment levels face less frequent price increases, they become relatively more attractive in subsequent rounds. Consequently, bidders willing to commit to faster deployment benefit from lower effective prices, aligning market incentives with policy goals of accelerating broadband coverage.

To estimate the cost of meeting deployment obligations, we integrate engineering assumptions with real infrastructure data. Our cost parameters draw on a U.S. FCC cost catalog originally compiled during the ZTE/Huawei infrastructure replacement program, which provides itemized estimates for tower construction, labour, legal, and fibre installation across 31 categories \cite{fcc2021scrp}. These base costs are adjusted to reflect Canadian market conditions: we apply a 10\% markup for U.S.–Canada market differences, a 10\% inflation factor, and a 30\% currency conversion premium. The resulting per-tower costs range from \$286,176 (low) to \$360,481 (high), before fibre backhaul is included.

Fibre costs are computed based on assumed separation distances between towers, varying by area type: 1km in metro/urban areas, 7.5km in rural zones, and 15km in remote regions. Fibre costs per kilometer are based on the FCC catalog and scaled using the same conversion parameters \cite{fcc2021scrp}. The total deployment cost is then computed by summing the tower installation cost and the estimated fibre length required to link towers together within each license area.

To determine how many towers a company would need to meet ISED’s deployment requirements, we use tower counts extracted from ISED’s Spectrum Management System (SMS) dataset as of December 31, 2024 \cite{ised2025smsdata}. We collapse the dataset’s radio-level entries into unique tower coordinates and assign each tower to a Tier 5 area. Each area is also classified as metro, urban, rural, or remote based on its Tier 5 designation. These counts are then aggregated up to the Tier 4 level (the license area granularity used in the auction) and merged with Statistics Canada population and land area statistics. Based on an assumed 20,000 population-per-tower ratio, we estimate the number of additional towers needed for each company to meet 5-, 10-, and 20-year deployment targets in every area.

However, tower density and fibre requirements vary by geography. To capture this, we introduce two forms of weighting:

\begin{itemize}
	\item \textbf{Population-based weighting} assumes that in denser areas, shorter fibre links are needed between towers (reflecting closer spacing), thus reducing cost.
	\item \textbf{Area-based weighting} assumes that larger, sparser regions require longer fibre links between towers, increasing cost.
\end{itemize}

While a comprehensive analysis would examine all combinations of cost levels and weighting schemes (a full 3x3 grid of 9 scenarios), we reduce computational overhead by selecting four representative scenarios. These are chosen to capture the key tradeoffs between cost intensity and geographic heterogeneity, as summarized in Table~\ref{tab:deployment_scenarios}.

\begin{table}%
    \caption{Deployment Cost Scenarios Used in the Extended Auction Simulation}
    \label{tab:deployment_scenarios}
    \begin{minipage}{\columnwidth}
        \scriptsize
        \begin{center}
            \begin{tabular}{p{4cm}p{9cm}}
                \toprule
                \textbf{Scenario} & \textbf{Cost Calculation Assumptions} \\
                \midrule
                No Cost Weighting\footnote{Tower cost uses the midpoint of low/high estimates. No population or area weighting applied.} &
                Mid-level deployment cost; uniform cost assumptions across all areas. \\

                Population-Weighted, High Cost\footnote{High cost base estimate (\$360,481); adjusted by population density to reduce fibre costs in denser areas.} &
                High deployment cost; fibre costs scaled by population-weighted multipliers to reflect urban/rural variation. \\

                Area-Weighted, Mid Cost\footnote{Mid deployment cost; fibre cost increased in larger geographic areas.} &
                Mid deployment cost; cost increases in large, sparse areas due to fibre length scaling. \\

                Combined Population and Area Weighting &
                Mid deployment cost; jointly weighted to reflect both population density and geographic size. \\
                \bottomrule
            \end{tabular}
        \end{center}
        \bigskip\centering
        \footnotesize\emph{Source:} ISED Spectrum Management System (Dec 2024), FCC Cost Catalog (adapted), Statistics Canada.

        \emph{Note:} Final deployment costs reflect both the number of new towers required and the estimated fibre needed to connect them.
    \end{minipage}
\end{table}%

The final deployment cost in each scenario consists of two components: (i) the number of towers required to meet the deployment obligation (based on the 5-, 10-, or 20-year threshold), and (ii) the estimated fibre cost needed to link these towers. For weighted scenarios (Area-Weighted and Combined Population and Area Weighting), this cost reflects demographic and spatial differences across Tier 4 license areas, accounting for variation in coverage challenges. Cost tables for each of these four scenarios are computed using the accompanying spreadsheet, with outputs organized by company and region in the “Scenario A” through “Scenario D” tabs.

It’s important to note that this cost model includes several simplifying assumptions. First, it assumes existing infrastructure can be reused optimally by each company, i.e., that current tower placements are already well-positioned for future use. It also assumes the cost of upgrading radio equipment is negligible relative to the cost of deploying new towers. Furthermore, while the model accounts for variation in population and land area, it does not account for topographical differences (e.g., mountainous regions), nor does it distinguish between different tower types or technologies beyond those already included in the SMS extract.

Despite these limitations, this deployment model provides a structured, data-driven foundation for simulating how deployment obligations and their associated costs, interact with bidder strategies in the extended auction. By embedding these differentiated costs into the product space, we enable a meaningful counterfactual analysis of how coverage incentives influence auction outcomes. 

We now turn to the results obtained from running our extended auction model under the four deployment cost scenarios described earlier. The extended auction (Auction $F$) was simulated for four major operators, Bell, TELUS, Rogers, Vidéotron, and Bragg, corresponding to the firms for which deployment costs could be reliably estimated using ISED’s Spectrum Management System tower data. To establish a benchmark, we also simulate the standard auction format (Auction $E$) with identical rules and companies but without deployment-level differentiation. In Auction $E$, a total of 3,787 licenses were sold at \$1,463,013,000 over 51 rounds.

\begin{table}[htbp]
	\centering
	\footnotesize
	\caption{Summary of Simulation Results Across Four Scenarios}
	\label{tab:scenario_results}
	\begin{tabular}{
			>{\raggedright\arraybackslash}p{2.5cm}
			>{\centering\arraybackslash}p{2.2cm}
			>{\centering\arraybackslash}p{2.2cm}
			>{\centering\arraybackslash}p{2.2cm}
			>{\centering\arraybackslash}p{2.8cm}
		}
		\toprule
		\textbf{Metric} & 
		\textbf{No Cost Weighting} & 
		\textbf{Population-Weighted} & 
		\textbf{Area-Weighted} & 
		\textbf{Combined Population and Area Weighting} \\
		\midrule
		Auction Revenue (\$) & 1,235,008,000 & 1,226,427,000 & 1,244,943,000 & 1,208,436,000 \\
		Low Licenses Sold & 2903 (67.51\%) & 2902 (67.49\%) & 2898 (67.40\%) & 2894 (67.30\%) \\
		Medium Licenses Sold & 365 (8.49\%) & 384 (8.93\%) & 377 (8.77\%) & 387 (9.00\%) \\
		High Licenses Sold & 488 (11.35\%) & 472 (10.98\%) & 472 (10.98\%) & 462 (10.74\%) \\
		\midrule
		Rural Coverage (Low/Med/High) & 1641 / 233 / 267 & 1641 / 252 / 251 & 1638 / 245 / 251 & 1634 / 245 / 251 \\
		Urban Coverage (Low/Med/High) & 1110 / 131 / 196 & 1109 / 131 / 196 & 1109 / 131 / 196 & 1109 / 141 / 186 \\
		Remote Coverage (Low/Med/High) & 152 / 1 / 25 & 152 / 1 / 25 & 151 / 1 / 25 & 151 / 1 / 25 \\
		\midrule
		Additional Population Covered & 3,331,585 & 3,331,585 & 3,331,585 & 3,331,585 \\
		\midrule
		Bell Cost (\$) for Higher Deployments & 42,104,946 & 44,705,629 & 54,721,791 & 54,721,791 \\
		Rogers Cost (\$) for Higher Deployments & 74,473,634 & 77,456,889 & 92,640,638 & 93,055,967 \\
		TELUS Cost (\$) for Higher Deployments & 0 & 0 & 0 & 0 \\
		Vidéotron Cost (\$) for Higher Deployments & 17,911,886 & 20,746,531 & 21,330,611 & 14,249,367 \\
		\bottomrule
	\end{tabular}
\end{table}

Across all four scenarios, the extended auction yields consistent outcomes that underscore both the robustness and policy relevance of the proposed mechanism. Total auction revenues range from \$1.21 to \$1.24 billion, an average of roughly 16\% below the baseline Auction $E$. This decline was predictable: the auction’s price-adjustment structure limits price escalation for higher deployment obligations, ensuring that their relative affordability reflects the additional investment required to meet those commitments. 

The resulting coverage improvements are substantial. Across all configurations, more than 3.3 million additional people would gain broadband access within five years relative to the non-extended design. This corresponds to increases of roughly 13\%, 10\%, and 9\% in remote, rural, and urban areas, respectively. These numbers are meaningful in the broader policy context: for example, ISED has set a national goal of increasing high-speed Internet access from 93.5\% in 2022 to 98\% by 2026, which is a roughly 4\% improvement in four years \citep{ised2025HighSpeedInternet}. It is also worth noting that the extended auction was simulated with only five participants. In a real-world auction with the full set of market competitors, competition would likely be stronger, and bidders may gravitate even more toward licenses associated with higher-level deployment obligations. In our simulations, approximately one-fifth of all licenses are allocated under medium or high deployment requirements. The stability of these results across varied cost assumptions indicates that the embedded incentive structure effectively aligns profit-maximizing bidding behavior with regulatory objectives, even in the presence of heterogeneous infrastructure costs.

On average, firms incur an additional \$152 million in deployment costs for medium and high tiers, suggesting that the burden of accelerated broadband rollout is shared, though unevenly, between the regulator and the private sector. TELUS, with its pre-existing infrastructure, incurs no additional cost, while Rogers and Bell bear the largest investment due to their broader geographic coverage. Vidéotron’s costs remain moderate, reflecting its regionally concentrated operations. Bragg is omitted from Table \ref{tab:scenario_results}, as it exclusively bids on low-deployment products, which are the only affordable options given its cost structure.

Overall, these results suggest that embedding deployment obligations directly into the auction format can meaningfully shift final allocations toward socially desirable outcomes. The mechanism enhances broadband coverage without compromising market efficiency. While short-term government revenue declines, the extended auction design supports long-term policy goals by accelerating infrastructure deployment and better aligning firm incentives with national connectivity goals.

\section{Conclusion}

This study demonstrates that a behaviorally grounded and computationally tractable model can effectively reproduce and analyze the complex dynamics of large-scale spectrum auctions. By estimating bidder valuations from observed round-by-round data and simulating myopic bidding behavior, we successfully replicated the key outcomes of Canada’s 3800 MHz auction with high fidelity, capturing allocation patterns almost exactly and revenue within a 6–10\% margin of the real event. The strong alignment between simulated and actual results validates the practical utility of simple, data-driven behavioral assumptions as an alternative to fully strategic equilibrium models.

Building upon this validated framework, we used the estimated valuations to conduct a counterfactual experiment incorporating deployment obligations directly into the auction design. The findings demonstrate that an integrated mechanism can materially improve broadband coverage, accelerating service for roughly 10\% of rural and remote populations (over 3.3 million individuals nationwide), while maintaining competitive integrity and moderate fiscal trade-offs. This finding underscores the potential of mechanism design informed by empirical valuation recovery: regulators can evaluate how modifications to auction rules would reshape incentives and outcomes without relying on intractable equilibrium computation.
Methodologically, our approach bridges the gap between structural econometric estimation and computational game theory. It offers a middle ground, anchored in observed data yet flexible enough for policy experimentation, enabling credible counterfactual analysis in environments where equilibrium assumptions are unrealistic. Despite its strengths, this work has several limitations that point to promising directions for future research. First, validating behavioral auction models remains inherently difficult, as public data typically come from a single observed auction. This raises the risk of overfitting to historical behavior rather than capturing general strategic principles. Developing new validation frameworks would strengthen the external validity of this approach. Second, the assumption of myopic bidders, while consistent with the structure, abstracts away from forward-looking strategic considerations that may influence real-world bidding. Extending the model to incorporate limited foresight, adaptive learning, or bounded rationality could yield a more nuanced representation of bidder behavior. Finally, the valuation estimates derived here could serve as an empirical foundation for subsequent equilibrium analysis. Using the recovered valuations to compute approximate Bayes–Nash equilibria or to evaluate the stability of observed outcomes would help connect empirical counterfactual analysis with theoretical predictions. Together, these extensions would not only address current methodological limitations but also enrich our understanding of strategic behavior and policy design in complex auction environments.


%
%
%
%
%

\bibliographystyle{ACM-Reference-Format}
\bibliography{Counterfactual_Analysis_of_Spectrum_Auctions_with_Application_to_Canada_s_3800MHz_Allocation}

\appendix

\end{document}